\documentstyle[epsfig,epsf,eqsecnum,aps,12pt]{revtex}
\def\beq{\begin{equation}}

\def\eeq{\end{equation}}

\def\bea{\arraycolsep .1em \begin{eqnarray}}

\def\eea{\end{eqnarray}}

\def\lsim{\mathrel{\lower4pt\hbox{$\sim$}}\hskip-12.5pt\raise1.6pt\hbox{$<$}\;}

\def\gsim{\mathrel{\lower4pt\hbox{$\sim$}}

\hskip-12.5pt\raise1.6pt\hbox{$>$}\;}

\def\s0#1#2{\mbox{\small{$ \frac{#1}{#2} $}}}

\def\0#1#2{\frac{#1}{#2}}

\begin{document}

\begin{center}

\thispagestyle{empty}

{\normalsize\begin{flushright}CERN-TH-99-400\\[12ex] 
\end{flushright}}

\mbox{\large \bf Masses of the Goldstone modes in the CFL  phase}\\[2ex]

\mbox{\large \bf  of QCD at finite density}\\[6ex]

{Cristina Manuel
\footnote{E-Mail:  Cristina.Manuel@cern.ch}${}$ and Michel H.G. Tytgat
\footnote{E-Mail:  Michel.Tytgat@cern.ch}${}$}
\\[4ex]
{\it  Theory Division, CERN, CH-1211 Geneva 23, Switzerland.}
\\[10ex]

{\small \bf Abstract}\\[2ex]

\begin{minipage}[t]{14cm}

We construct the $U_L(3) \times U_R(3)$ effective lagrangian which encodes
the dynamics of the low energy pseudoscalar excitations
in the Color-Flavor-Locking superconducting phase of QCD  at finite quark density.
We include the effects of instanton-induced interactions and study the
mass  pattern of the pseudoscalar mesons. A tentative comparison with  the  
analytical estimate for the  gap  suggests that some of these low energy momentum modes
are not stable for moderate  values of the quark chemical potential $\mu$.

\end{minipage}

\end{center}

\newpage

\pagestyle{plain}

\setcounter{page}{1}


\section{Introduction} 


The fascinating possibility that the ground state of  QCD at finite quark 
density 
can behave as a color superconductor has attracted much attention recently
\cite
{Bailin:1984,Alford:1998,Rapp:1998,Alford:1999,Schafer:1999,Alford:1999b,Kogut:1999,Evans,Evans:1999at,Hsu:1999mp,Son:1999,Schafer:1999b,Pisarski:1999,Hong:1998,Schafer:1999fe,Rapp:1999}.
Of particular interest is the symmetry breaking pattern
that may arise for $N_f=3$ light quark flavors. 
As first shown by Alford, Rajagopal and Wilczek \cite{Alford:1999},  
the diquark condensates can lock the color and flavor symmetry transformations 
(Color-Flavor-Locking or CFL for brief). In the limit of massless quarks, these condensates
spontaneously break both color and chiral symmetries. The eight gluons 
become massive through
the Higgs mechanism while eight plus two  
Goldstone bosons are leftover. The two octets of states are analogous
to the octets of vector and  light pseudoscalar mesons in vacuum and Sch\"afer and Wilczek \cite{Schafer:1999}
have conjectured that there might exist some sort of continuity 
 between the properties of QCD at zero density and in the CFL phase.
The two other massless excitations are 
the Goldstone modes associated with 
the spontaneous breaking of the baryon number ($U(1)_B$) and axial ($U(1)_A$) symmetries. 
Because of the axial anomaly, the latter is not a true symmetry of QCD.
However, because instantons effects are small at large densities,
the associated mode can be treated as a true Goldstone mode as a first 
approximation.
Although there are significant differences, to which we shall come back later, the situation at high density
is analogous to considering the limit of large numbers of colors $N_c$ in vacuum \cite{largeN}, 
in which  $U(1)_A$ breaking effects are $1/N_c$ suppressed.

For energies which are small compared to the gap of the superconducting CFL phase, the dynamics of these Goldstone modes
is most conveniently described with the help of an 
effective lagrangian. 
As suggested by the symmetry breaking pattern in the CFL phase at large densities, this effective theory
is analogous to Chiral Perturbation Theory ($\chi$PT) in the large $N_c$ limit of  QCD in vacuum \cite{Gasser:1985}.
For  large densities, the leading
order effective lagrangian invariant under $U(3)_L \times U(3)_R$ flavor symmetry has been constructed in 
some recent works~\cite{Hong:1999dk,Casalbuoni:1999wu,Casalbuoni:1999zi,Son:1999cm,Hong:1999ei,Rho:2000ww}. 
In particular, Son and Stephanov \cite{Son:1999cm} have shown how the parameters of the lagrangian
can be computed  at large densities
by matching to the underlying microscopic theory. 
In the present note we make a first attempt to  extend these works
to lower density regimes, taking into account the effects of instanton-induced interactions.
By using the power counting rules of $\chi$PT we construct the effective lagrangians
up to order $E^4$ in an energy expansion. 
We  make a particular emphasis on the meson spectrum and compute their masses as function of the quark chemical potential $\mu$.
For moderate densities, $\mu \lsim 10^4$ MeV, the gauge coupling grows large and instanton effects become 
important. We give numerical values for the masses of the mesons simply assuming that the analytical expressions for the
gap and condensates, which have been  computed at weak gauge coupling, also hold at strong coupling. 
Our calculations suggests that some of the Goldstone modes cannot exist
as low energy excitations for values of $\mu < 1000$ MeV,
as their masses are very close to the
unstability threshold $2 \Delta$, where $\Delta$ is the gap in the CFL phase.
These conclusions depend on the value of the color superconducting gap
as estimated in the literature, and of a number of 
approximations that we have made.

In the next section, we write down the general low energy effective action for
the pseudoscalar Goldstone modes up to $E^4$ in a low energy expansion and estimate the couplings
that are  relevant for the meson mass spectrum. In Sec. \ref{sec.numb} we give 
some numbers for the meson masses at low densities, $\mu \lsim 2500$ MeV and finally draw some conclusions.


%

\section{Effective chiral lagrangian in the CFL phase} 

\label{ECHPL}


We follow  Gatto and Casualboni \cite{Casalbuoni:1999wu} and  Son and Stephanov \cite{Son:1999cm} to construct 
the effective lagrangian for
the low energy excitations of the CLF phase of QCD. 
The ground state is characterized by the two diquark condensates

\begin{equation}
\label{condensates}
X^{ia} \sim \epsilon^{ijk} \epsilon^{abc} \langle \psi^{bj}_L \psi^{ck }_L \rangle^* \ , 
\qquad  Y^{ia} \sim \epsilon^{ijk} \epsilon^{abc} \langle \psi^{bj}_R \psi^{ck }_R \rangle^* \ ,
\end{equation}
where  $a,b,c$ denote color indices, while $i,j,k$ refer to flavor ones. 
Under an $SU(3)_c \times SU(3)_L \times SU(3)_R$ the condensates transform as
\begin{equation}
X \rightarrow  U_L X U_c^{\dagger} \ , \qquad Y \rightarrow  U_R Y U_c^{\dagger} \ ,
\end{equation}
and as
\begin{equation}
X \rightarrow e^{2 i \alpha} \, e^{2 i \beta} X \ , \qquad Y \rightarrow  e^{- 2 i \alpha} 
\, e^{2 i \beta} Y \ ,
\end{equation}
under $U(1)_A$ and $U(1)_B$ transformations defined as
\begin{equation}
\psi_L \rightarrow e^{i(\alpha + \beta)} \psi_L\ , \qquad \psi_R \rightarrow e^{i(-\alpha + \beta)} \psi_R
\ .
\end{equation}
One can factor out the norm of the condensates and consider
the unitary matrices $X$ and $Y$. The 
slow variations of the phases of these matrices then correspond to the low energy excitations. 
Altogether these are $9+9 =18$ degrees of freedom: $8$ will be absorbed by the gluons through the 
Higgs mechanism, which leaves $10$ true low energy excitations. At low energies the gluons
decouple from the theory, as they are heavy degrees of freedom. It is convenient to collect all the
Goldstone modes in the unitary matrix
\begin{equation}
\label{umatrix}
\Sigma = X Y^\dagger \ ,
\end{equation}
which is singlet of $SU(3)_c$ and $U(1)_B$ and transforms as
\begin{equation}
\Sigma \rightarrow e^{4 i \alpha} \,U_L \Sigma U_R^\dagger \ ,
\end{equation}
under $SU(3)_L \times SU(3)_R \times U(1)_A$. This leaves apart the low energy excitation that emerges from the spontaneous breaking of
baryon number symmetry $U(1)_B$. Because baryon number is an exact global symmetry of QCD, this Goldstone mode is 
always massless in the CFL phase, independent of the quark masses. 
As our focus here is on the effect of chiral symmetry breaking by finite quark
masses, to simplify our discussion  we will simply drop this degree of freedom in the sequel.

To proceed we  parametrise the unitary matrix $\Sigma$ as  
\begin{equation}
\Sigma =  \exp \left(i\,{\Phi\over f_\pi} \right) \ ,
\end{equation}
where $\Phi = \phi^A T^A\ $, $A=1,\ldots 9$. The $T^A$ for $A=1, \ldots, 8$
are the Gell-Mann generators of $SU(3)$ 
and $T^9 = \sqrt{2/3}\,{\bf 1}_3$, all normalized 
to ${\rm Tr} (T^A T^B) = 2 \delta^{AB}$. We use the same 
nomenclature as in vacuum for the nonet of
Goldstone modes,
\begin{equation}
\label{names}
\Phi =  \,\left(
\begin{array}{ccc}
 \pi_0 + {1\over \sqrt 3} \left( \eta_8 
 + \sqrt{2} { f_\pi\over f_{\eta_0}} \eta_0 \right) & \sqrt{2} \pi^+ &\sqrt{2} K^+ \\
\sqrt{2} \pi^- & - \pi_0 + {1\over \sqrt 3} \left( \eta_8 + \sqrt {2} { f_\pi\over f_{\eta_0}} \eta_0 \right)& \sqrt{2} K_0\\
\sqrt{2} K^- & \sqrt{2} \bar K_0 &  {1\over \sqrt 3 } \left(-2 \eta_8 + \sqrt{ 2} { f_\pi \over f_{\eta_0} } \eta_0 \right)\\ 
\end{array}\right) \ .
\end{equation}
In (\ref{names}), $f_\pi$ and $f_{\eta_0}$ are the decay 
constants at finite density respectively
of the octet of  mesons and of the singlet  $\eta^0$. 
{\em A priori} there is no reason for these to be equal.
By matching to the microscopic theory, Son and Stephanov \cite{Son:1999cm} have found
\begin{equation}
\label{deconst}
f_\pi^2 = {21 - 8 \log 2\over 18} \, {\mu^2\over 2 \pi^2} \ ,  \qquad
f_{\eta_0}^2 = {3\over 4}\, {\mu^2\over 2 \pi^2} \ ,
\end{equation}
where $\mu$ is the quark chemical potential. These 
expressions are valid at large densities ({\em i.e.} small gauge coupling). From~(\ref{deconst}), the ratio $f_\pi/f_{\eta_0} 
\approx 1.07$ is close to one. This result is reminiscent of
the $OZI$ rule often invoked to hold in vacuum. At smaller densities, the gauge coupling grows large
and the ratio could significantly depart from unity because of instanton effects.

The quark mass term in the microscopic lagrangian
\begin{equation}
\Delta {\cal L} = - \bar\psi_L  {\cal M} \psi_R + h.c.
\end{equation}
breaks chiral symmetry. Its effects can be introduced in the 
effective lagrangian  by treating the mass matrix as an external 
field with vacuum expectation value
\begin{equation}
{\cal M} = \mbox{diag}(m_u,m_d,m_s) \ ,
\end{equation}
where $m_u, m_d$ and $m_s$ refer to the up, down and strange quark masses,  respectively.
The spurion field then transforms  as 
\begin{equation}
{\cal M} \rightarrow  e^{-2 i \alpha} U^\dagger_L {\cal M}U_R \ ,
\end{equation}
under $SU(3)_L \times SU(3)_R\times U(1)_A$.

To take into account the effects of the $U(1)_A$ anomaly, we also allow for
the presence of the $\theta$ term in the microscopic lagrangian,
\begin{equation}
\Delta{\cal L}_\theta = \theta \frac{g^2}{32 \pi^2} F_{\mu \nu}^A {\tilde F}^{\mu \nu, A} \ .
\end{equation}
As for the quark mass term, we will treat
$\theta$ as an external field with vanishing expectation value  
which, to account for the variation of the quark measure,
transforms as
\begin{equation}
\theta \rightarrow \theta - 2 N_f \alpha \equiv  \theta - 6\, \alpha
\end{equation}
under $U(1)_A$ transformations.  Then any arbitrary function of the combination
\begin{equation}
\label{comb}
X=  \theta - \frac{i}{2} {\rm Tr} \log{\Sigma }  \equiv  \theta + \sqrt{3\over 2} {\eta_0 \over f_{\eta_0}}  ,
\end{equation}
is invariant under $SU(3)_L \times SU(3)_R \times U(1)_A$ transformations.

\bigskip

With these ingredients, we are almost ready to
construct the low energy effective lagrangian. One last issue is that of power counting. 
In QCD in  vacuum, because $M_\pi^2 \sim m_q$, the expansion is in powers of the pion  external energy or  momenta and quark masses and
$E^2 \sim p^2 \sim m_q$. 
Similarly, at large $N_c$,   $U(1)_A$ breaking effects are
counted as $E^2 \sim 1/N_c$, because $M_{\eta_0}^2 \sim 1/N_c$.  
In the CFL phase of QCD on the other hand, the power counting depends very 
much on the density. At finite densities,  instantons effects are screened
$\propto (\Lambda/\mu)^\alpha$, where $\Lambda \approx 200$ MeV 
is the scale of QCD, $\mu$ is the quark chemical potential and
$\alpha$ is some positive constant which
depends on the process under consideration.
This has two consequences at  large densities. The first is that $U(1)_A$ is essentially a good symmetry
and the $\eta_0$ is approximately massless in the chiral limit. 
Then,  because the two condensates break chiral symmetry only
through the intermediate of color
transformations, there is an approximate $Z_2^L \times Z_2^R$ symmetry
which acts independently on the left and right-handed
quarks~\cite{Alford:1999}.  At high densities, this implies that 
the leading contribution to the mass of the Goldstone modes is  ${\cal O}(m_q^2)$. The natural power counting rule at high densities is then $E^2 \sim m_q^2$ because $M_\pi^2 \sim m_q^2$. At the
moderate densities that could eventually be of interest for heavy ion
collisions or for neutron stars,
instantons effects are likely to be non-negligible. As $Z_2^L\times Z_2^R$ symmetry is broken to the
diagonal $Z_2$ by instantons,  a meson mass 
term $M_\pi^2 \sim m_q$ is allowed \cite{Alford:1999}. If instanton effects are dominant, the power counting is
that of vacuum, $E^2 \sim m_q$. For convenience, we will adopt the 
counting rules of $\chi$PT  as in vacuum
and comment when necessary about the differences
that arise in the CFL phase of QCD.

A final remark concerning the region of applicability of the energy expansion. 
In vacuum, the low energy expansion is valid for
$E^2 \lsim f_\pi^2$. In the CFL phase however, $f_\pi \sim \mu$, which is independent and much larger than  
the gap ($\Delta \ll \mu$) at large densities (see~(\ref{deconst})). However, the effective theory must break down for
$E \sim 2 \Delta$, which is the energy necessary to excite one quasiparticle pair out of the superconducting 
ground state, and the low energy expansion is only valid for the more 
restrictive range 
$E^2 \sim M_\pi^2 \lsim 4 \Delta^2$~\footnote{There is a  formal
analogy between this behavior and that of $\chi$PT in vacuum 
in the large $N_c$ limit. {\em A priori} the expansion is valid for
$E < f_\pi$. But because
$f_\pi^2 \sim N_c \Lambda^2$   grows large  as $N_c\rightarrow \infty$ and
as  something is bound to happen for
$E \sim \Lambda \sim 200$ MeV, which is the scale of quark confinement, the low  energy expansion is 
limited to  $ E \lsim  \Lambda$.}.

\bigskip

At order $E^2$, the most general lagrangian compatible with the symmetries
of the condensates is
\begin{eqnarray}
\label{Etwo}
{\cal L}_2 &= &  V_1(X) {\rm Tr} \left( \partial_\mu \Sigma
\partial^\mu \Sigma^\dagger \right) +
\left\{ V_2 (X) {\rm Tr} \left( {\cal M} \Sigma^\dagger \right) e^{i
\theta} + h. c. \right\}\nonumber \\
&+& V_3(X) \left({\rm Tr} \,\Sigma \partial_\mu \Sigma^\dagger \right)^2 + 
 V_4(X)  \left({\rm Tr} \,\Sigma \partial_\mu \Sigma^\dagger \right) \partial^\mu \theta
+V_5(X) \partial_\mu \theta \partial^\mu \theta \ .
\end{eqnarray}
We have written (\ref{Etwo}) using a compact notation.
Since Lorentz invariance
is broken at finite density (only a $O(3)$ symmetry is preserved)
all the four-vectors should be split into temporal and
spatial components, and the functions $V_i$ multiplying those spatial or
temporal components  are not forced to be the same by symmetry arguments.
For example, the first term in (\ref{Etwo}) should read
\begin{equation}
V_1 (X) {\rm Tr} \left( \partial_\mu \Sigma
\partial^\mu \Sigma^\dagger \right) \rightarrow 
V_{1, t} (X) {\rm Tr} \left( \partial_0 \Sigma
\partial_0 \Sigma^\dagger \right) - V_{1,s} (X)  {\rm Tr} \left( \partial_i
\Sigma \partial_i  \Sigma^\dagger \right) \ .
\end{equation}
The same situation occurs for the remaining terms and functions $ V_3, V_4, V_5$ of $X$.
All the couplings in (\ref{Etwo}) can  be {\em a priori}  arbitrary functions of $X$ (\ref{comb}). 
At large densities, $U(1)_A$ symmetry breaking effects are exponentially suppressed and the couplings only depend on the chemical potential $\mu$.

To reproduce the standard normalization of the meson kinetic terms, we impose
\begin{equation}
V_{1,t}(0) = \frac{f^2_\pi}{4} \ .
\end{equation}
The ratio $V_{1,s}(0)/V_{1,t}(0) = v^2$ 
is the  velocity squared of the 
Goldstone bosons.  At large densities, Son and Stephanov \cite{Son:1999cm} have found that $v$ is equal to the
speed of sound $1/\sqrt{3}$ for  all the low energy modes, including the baryon Goldstone mode.

The operators in (\ref{Etwo}) are not all 
independent. The last three terms can be transformed into each other with 
a field redefinition, $\eta_0/f_{\eta_0} \rightarrow \eta_0/f_{\eta_0} + \kappa \theta$. Using this freedom, we can
choose to set $V_4(0)$ to zero. With this choice, 
the last operator becomes irrelevant for the meson spectrum 
and can be discarded. 

The operator that is left, with coupling $V_3(X)$, contributes to the difference between $f_\pi$ and $f_{\eta_0}$. 
At high densities, using~(\ref{deconst}),  
\begin{equation}
V_{3,t}(0) = {f_\pi^2 - f_{\eta_0}^2\over 12} \approx 0.01 f_\pi^2,
\end{equation}
which is small compared to $f_\pi^2$. At moderate densities, $V_{3,t}$ could receive large contributions 
from instantons as is manifest from the mixing with $V_4$ and $V_5$.

Consider now the mass term in (\ref{Etwo}).
This term is analogous to the leading mass term in
$\chi$PT in vacuum. The only difference is the occurrence of the phase 
$\theta$,
 which is absent in vacuum. This is  because at zero density the  condensate that breaks chiral symmetry is 
$\langle \bar \psi_L \psi_R\rangle$ which 
transforms like ${\cal M}^\dagger$  under $SU(3)_L \times SU(3)_R \times U(1)_A$.  In (\ref{Etwo}) the presence of the $\theta$ in the effective lagrangian is
the trademark of
a one instanton effect; in the CFL phase, a one instanton process can be saturated by closing its six external quark legs
with the insertion of one left-handed diquark, one right-handed diquark and one chiral condensate, 
$\sim (\psi_L \psi_L) (\bar\psi_R\bar\psi_R) (\bar\psi_R\psi_L)$. 
As in vacuum, a non-zero chiral condensate $\langle \bar\psi\psi\rangle$ leads to $M_\pi^2 \propto m_q \langle \bar \psi \psi \rangle$. 
Instanton effects
are small and can be reliably computed at high densities $\mu \gg \Lambda$.
In particular,  Sch\" afer \cite{Schafer:1999fe} has obtained
\begin{eqnarray}
\label{inst.cond}
\langle \bar \psi \psi \rangle &= & \langle \bar u u + \bar d d + \bar s s \rangle \nonumber \\
&\approx &  
- 2 \left ( {\mu^2 \over 2 \pi^2}\right ) {3 \sqrt{2} \pi\over g(\mu)}\,{18\over 5} G(\mu) \phi_A^2(\mu) \ ,
\end{eqnarray}
where $g(\mu)$ is the gauge coupling estimated at
the scale $\mu$ using the one-loop beta function. The factor $G(\mu)$ 
is the one instanton weight integrated over all instanton
sizes $\rho$, which  are peaked around $\rho \sim \mu^{-1}$ at finite density, 
\begin{equation}
\label{Gfunct}
G(\mu) \approx 
0.26 \, \Lambda^{-5}\, \left((\beta_0 \log(\mu/\Lambda)\right)^{2 N_c} \left 
( {\Lambda\over \mu}\right  )^{\beta_0 +5 }.
\end{equation}
We have used the running of $g(\mu)$ at 
one-loop, $\Lambda = 200 $ MeV 
and  $\beta_0 = 11/3 \,N_c - 2/3\, N_f \equiv 9$ is the first coefficient of the QCD beta function. 
Finally, $\phi_A(\mu) \sim \langle\psi_L \psi_L\rangle \sim \langle\psi_R\psi_R\rangle$ is the diquark condensate in the CFL phase of QCD, 
\begin{equation}
\phi_A(\mu) \approx 2 \left 
( {\mu^2\over 2 \pi^2 } \right ) \left({3 \sqrt{2} \pi\over g(\mu)}\right)\, \Delta(\mu) \ ,
\end{equation}
and \cite{Son:1999,Schafer:1999,Pisarski:1999}
\begin{equation}
\label{colorgap}
\Delta (\mu) = b_0^\prime 512 \pi^4 (2/N_f)^{5/2} \, g(\mu)^{-5}\, \mu\, \exp\left ( -{3 \pi^2\over \sqrt{2}  g(\mu) }\right ) \ ,
\end{equation}
is the color superconducting gap. The factor $b_0^\prime$ is unknown but expected to be ${\cal O}(1)$ ~\cite{Pisarski:1999,Brown:1999aq,Brown:1999yd}.
At large  densities 
$\mu \gg \Lambda$, instantons are suppressed, $\langle \bar \psi \psi\rangle$ goes to zero and the mass term
of  (\ref{Etwo}) vanishes.  At the moderate densities
that could be of interest for heavy ion collisions or neutron stars, this instanton effect is however not negligible. 
Using  the Gell-Mann-Oakes-Renner relation, and 
$\langle \bar \psi \psi\rangle = \langle \bar u u + \bar d d + \bar s s\rangle \equiv 3 \langle \bar u u \rangle$, we find
\begin{equation}
\label{vtwo}
V_2(0) =  - \frac{1}{6} \langle \bar \psi \psi\rangle  \ .
\end{equation}

\bigskip

At order $E^4$, there are a few more operators\footnote{At this order, one could also consider adding a Wess-Zumino-Witten term~\cite{Hong:1999dk}.},
\begin{eqnarray}
\label{Efour}
{\cal L}_4 &= &  K_1(X)\left[ {\rm Tr} \left( \partial_\mu \Sigma
\partial^\mu \Sigma^\dagger \right)\right]^2 + K_2(X){\rm Tr} \left( \partial_\mu \Sigma
\partial_\nu \Sigma^\dagger \right) {\rm Tr} \left( \partial^\mu \Sigma
\partial^\nu \Sigma^\dagger \right) \\
& + &K_3(X) {\rm Tr} \left( \partial_\mu \Sigma
\partial^\mu \Sigma^\dagger \partial_\nu \Sigma
\partial^\nu \Sigma^\dagger\right) \nonumber \\
&+& \left\{ K_4(X) {\rm Tr} \left( \partial_\mu \Sigma
\partial^\mu \Sigma^\dagger \right) {\rm Tr} \left( {\cal M} \Sigma^\dagger \right) e^{i
\theta} + h. c. \right\}+ \left\{ K_5(X) {\rm Tr} \left( \partial_\mu \Sigma
\partial^\mu \Sigma^\dagger {\cal M} \Sigma^\dagger\right)  e^{i \theta} +h.c.\right\} \nonumber \\
&+& \left\{ K_6(X) \det(\Sigma) \, {\rm Tr} \left(
 {\cal M} \Sigma^\dagger \right){\rm Tr} \left( {\cal M} \Sigma^\dagger \right)
+ h.c. \right\}+ K_7(X)  {\rm Tr}\left( {\cal M} \Sigma^\dagger \right)
{\rm Tr} \left( {\cal M}^\dagger \Sigma \right) \nonumber \\
&+& \left\{ K_8(X) \det(\Sigma) \, {\rm Tr} \left( {\cal M} \Sigma^\dagger 
{\cal M} \Sigma^\dagger \right)
+ h.c. \right\} \ .
\nonumber
\end{eqnarray}
Again, we have used 
a compact notation for the terms in $K_i$, $i=1\ldots 5$ which should be split into temporal and spatial components.

Like the term $V_2$ in~(\ref{Etwo}), 
the terms $K_4$ and $K_5$ depend explicitly on $\theta$ and so are exponentially suppressed at high densities.
Similarly, at  large chemical potential $\mu$, the remaining functions  $K_i(X)$ reduce to constants $K_i(X=0;\mu)$. 
In this limit, the mass pattern of the Goldstone bosons
will be determined  by the couplings $K_6$, $K_7$ and $K_8$ in (\ref{Efour}). 
As first shown by Son and Stephanov \cite{Son:1999cm}, at large densities, these coupling constants can be computed by 
matching to the underlying microscopic theory.

\bigskip

In principle,  a systematic and non-ambiguous strategy
to compute the coefficients of the effective theory is to use the background field technique \cite{Espriu:1990},  introducing external 
sources and symmetry breaking order parameters and   integrating
out the quark fields. At high densities,  gluon exchange is suppressed and this amounts to a  
quark one-loop calculation, which 
would fix the coefficients of all the operators to arbitrary order in the
meson fields. A slightly different strategy  has been advocated in \cite{Son:1999cm}: set the meson fields to zero, 
$\Sigma = {\bf 1}_3$ in the operators of (\ref{Efour}), and compare 
the shift in ground state energy induced by non-zero quark masses in both
the effective and microscopic theories. 
This approach is {\em a priori} ambiguous, as different 
operators could contribute to the shift in ground state energy. In practice,
 to  order $E^4$, enough constraints
can be derived to completely fix the couplings $K_6$, $K_7$ and $K_8$. 
The diagrams in the full microscopic theory are those of Fig.1. The {\em rhs} 
diagram can only contribute to $K_6$ and $K_8$.
The {\em lhs} diagram can contribute to $K_7$. 
These couplings can be fixed by considering two different quark mass patterns, 
\begin{equation}
{\cal M}_1 = m {\bf 1}_3\;\;\; \mbox{and} \;\;\; {\cal M}_2 = {\rm diag}(0,0,m_s) = - 
{m\over\sqrt{3}} \, \lambda^8 + {m\over\sqrt{6}}\, \lambda^9 \ .
\end{equation}
In the effective theory, these respective choices lead to the following shifts in ground state energy density
\begin{eqnarray}
\Delta \varepsilon_1 &=&- \left\{ (9 m^2 K_6 + h.c.) + 9 m^2 K_7 + (3 m^2 K_8 + h.c.)\right\}  \ ,
\\
\Delta \varepsilon_2 &=&-\left\{ ( m_s^2 K_6 + h.c.) +  m_s^2 K_7 + (m_s^2 K_8 + h.c.) \right\} \ .
\end{eqnarray}

A special feature of a quark mass term is that it couples fermions and antifermions. At finite density, quark mass effects
are then suppressed by the necessity to excite antiparticles with characteristic momentum $\sim 2 \mu$~\cite{Hong:1999ei,Rho:2000ww}.
A simple way to see this is to consider the effective theory for the fermion
excitations near the surface of the Fermi sea. 
In momentum space, the free fermion lagrangian is
\begin{equation}
{\cal L} = \bar \psi (\gamma^\mu p_\mu + \mu \gamma^0) \psi - m \bar \psi 
\psi \ .
\end{equation}
If we consider the fermion mass term as a 
perturbation and decompose the 
fermions field $\psi$ into positive and negative energy components $\psi_+$ and $\psi_-$, using the projectors
\begin{equation}
\Lambda^\pm = {1\over 2} (1 \pm {\bf \alpha}\cdot {\bf \hat q}) \ ,
\end{equation}
and $\Lambda^\pm \psi_\pm = \psi_\pm$, the lagrangian becomes
\begin{equation}
{\cal L} = \psi_+^\dagger (q_0 - \vert \vec q \vert + \mu)\psi_+ + \psi_-^\dagger (q_0 + \vert \vec q \vert + \mu)\psi_- 
- m (\psi_+^\dagger \psi_- + \psi^\dagger_- \psi_+) \ .
\end{equation}
This shows that the fermion mass term couples particles and antiparticles. 
If we integrate out the antifermions, we get at leading order
\begin{equation}
{\cal L} \approx \psi_+^\dagger (q_0 -  q_\| )\psi_+ - {m^2\over \mu} \psi^\dagger_+ \psi_+
\end{equation}
where $q_\| \approx \vert \vec q \vert - \mu$. This shows that quark mass 
effects are suppressed at high densities, as one could have expected.
Note that the modified mass term $\propto m^2/\mu$ has the same structure as a chemical potential term. This is a
relevant operator which modifies the shape of the Fermi surface~\cite{Polchinski:1992ed}.

\bigskip

The calculation of the {\em l.h.s.} diagram of Fig. 1 gives 
\begin{equation}
{\Delta \varepsilon_1\over \Delta \varepsilon_2} = 3 {m^2\over m_s^2} \ ,
\end{equation}
to ${\cal O}({m_q^2 \Delta^4\over \mu^4})$, where $\Delta$ stands for generic 
gap and/or antigaps.
This diagram simply gives a trivial  shift of the 
vacuum energy, which in the effective theory corresponds to a constant (independent of $\Sigma$) invariant  operator
\begin{equation}
\Delta {\cal L} \propto \mbox{Tr}({\cal M}^\dagger {\cal M}) \ ,
\end{equation}
and, consequently, $K_7=0$.

Estimates of the {\em r.h.s.} diagram of Fig. 1
indicate that~\cite{Hong:1999ei}  
\begin{equation} 
\label{ksix}
K_{6,8}  \sim  {\Delta \bar \Delta \over \mu^2} \log (\Delta/\mu) \ ,
\end{equation}
where $\Delta$ is a generic expression for a quark gap and $\bar \Delta$ is an antigap. The dependence on antigaps arises 
from the structure of the {\em r.h.s.} diagram with both particles and antiparticles propagators. Unfortunately, the expression of the antigap is not 
known. Preliminary estimates have revealed that it is gauge-dependent \cite{Schafer:1999b}. It is however a physical quantity 
(pole of the antiquark quasi-particles and holes), which should be gauge-invariant on quasi-particles mass-shell. 
At weak coupling, the antigap is presumably much smaller than the 
gap. Here, we will assume that $K_{6,8} \approx 0$ at large and moderate 
densities.

\bigskip

To be complete we should also include the contribution of instantons to the
mass of the
singlet meson $\eta_0$,
which enters the effective lagrangian through a function of $X$ alone,
\begin{equation}
\label{Ezero}
{\cal L}_0 = - V_0 (X) \ ,
\end{equation}
This piece  is $E^0$ as it involves no derivatives of 
the meson fields. Expanding to second order in $X$ gives
\begin{equation}
M_{\eta_0}^2 = {3\over 2} \,{V_0^{''}(0)\over f_{\eta_0}^2} \ .
\end{equation}
The constant $V_0^{''}$ (and more generally the whole function $V(X)$)  could in principle be computed at large densities using instanton calculus.  
Unfortunately, the result is not known. However, at large to moderate 
densities we might expect this contribution to  be suppressed. 

For three flavors, 
the two remaining legs of a one instanton contribution to $M_{\eta_0}^2$ can only be closed with the insertion of either a quark mass term or a chiral 
condensate $\langle \bar \psi \psi\rangle$. In the chiral
limit, the latter only arises through another instanton process. 
On dimensional grounds we would  expect
\begin{equation}
M_{\eta_0}^2 \propto \mu^4 \, G(\mu) \vert \langle  \bar \psi \psi \rangle\vert
\end{equation}
in the chiral limit,
but from such  a rough estimate, it is not reasonable 
to infer whether $M_{\eta_0}$ ever grows large at low densities.


%

\section{Meson mass pattern}

\label{sec.numb}


In this section we give  numerical estimates of the pseudoscalar mesons
as  function of the quark  chemical potential.

At very large densities, instanton effects 
are suppressed:  meson masses are linear in the quark masses and presumably
very small. At lower densities, 
a non-zero $\langle \bar \psi \psi \rangle$ condensate induced by instantons 
introduce  contributions to the meson masses which are proportional to
$m_q^{1/2}$. Here, we will  {\em  estimate} this last effect 
as it is dominant in the low density regime of the theory. We will  
use the expression of the chiral condensate and the gap as computed at weak 
coupling (\ref{inst.cond}) and we will assume that the corrections
are small even  at moderate densities $\mu \sim 600$ MeV.

The masses of the charged pions and kaons deduced from (\ref{Etwo}) read 
\begin{equation}
\label{othermasses}
M^2_{\pi^\pm} =   {2 V_2\over f_\pi^2}  \, \left(m_u + m_d \right ) 
\ , \qquad 
M^2 _{K^\pm}  =    {2 V_2\over f_\pi^2}  \, \left(m_u + m_s \right )
 \ , \qquad
M^2 _{K^0, {\bar K}^0}  =   {2 V_2\over f_\pi^2}   \, \left(m_d + m_s \right )   \ ,
\end{equation}
In the limit of no $U(1)_A$ breaking effects, the neutral mesons $\pi^0, \eta_8$ and $\eta_0$ are strongly mixed.
Their mass matrix reads
\begin{equation}
\label{neutralmass}
{2 V_2 \over f^2_\pi}  \,\left(
\begin{array}{ccc}
m_u + m_d & \frac { m_u -m_d}{\sqrt{3}} & 
\frac{f_\pi}{f_{\eta_0}} \, \frac{ \sqrt{2} (m_u -m_d)}{\sqrt{3}} \\
\frac {2 (m_u -m_d)}{\sqrt{3}} & \frac{m_u + m_d + 4 m_s}{3} &
\frac{f_\pi}{f_{\eta_0}} \, \frac{\sqrt{2} (m_u +m_d -2m_s)}{  \sqrt{3}} \\
\frac{f_\pi}{f_{\eta_0}} \, \frac{ \sqrt{2} (m_u -m_d)}{\sqrt{3}} &
\frac{f_\pi}{f_{\eta_0}} \, \frac{\sqrt{2} (m_u +m_d -2m_s)}{\sqrt{3}}&
\frac{f^2_\pi}{f^2_{\eta_0}} \, \frac{ 2 (m_u +m_d +m_s)}{ 3 } 
\end{array}\right) \ .
\end{equation} 
We have  assumed that the contribution from the quark masses and
the chiral condensate
is parametrically larger than the two-instantons contribution to the mass of
$\eta_0$.
As the latter effect, if important, would  increase the mass of $\eta_0$,
our numerical estimates
should be considered as lower bounds on the mass of $\eta^\prime$.   
Neglecting the difference between $f_\pi$ and $f_{\eta_0}$ we get the
mass eigenvalues characteristic of ideal mixing,
\begin{equation}
M^2_{\bar u u} =  {2 V_2\over f_\pi^2}  \, 2 m_u \ , \qquad
M^2_{\bar d d} =  {2 V_2\over f_\pi^2}  \, 2 m_d \ , \qquad
M^2_{\bar s s} =  {2 V_2\over f_\pi^2}  \, 2 m_s \ .  
\end{equation}
To give an idea of the numerical values of the meson masses
at $\mu \sim 600$ MeV, 
we take $m_u = 4$ MeV,  $m_d = 7$ MeV and $m_s = 150$ and evaluate
the value of the chiral condensate for $\Lambda = 200$ MeV and $b'_0 = 1$,
to find
\begin{eqnarray}
M^2_{\pi^\pm}  \approx    30\, {\rm MeV} 
 \ , \qquad 
 M^2 _{K^\pm}  \approx 112\, {\rm MeV}
 \ , \qquad
M^2 _{K^0, {\bar K}^0}  \approx 113\, {\rm MeV}    \ , & \\
M^2_{\bar u u} \approx 26\, {\rm MeV} , \qquad
M^2_{\bar d d} \approx 34\, {\rm MeV} \ , \qquad
M^2_{\bar s s} \approx 71\, {\rm MeV} \ . &  
\end{eqnarray}
A more accurate analysis of the meson masses would require to take
into account terms which  are proportional to $m_q^2$.
However, in the low density regime, and due to
the  smallness of $K_6$ and $K_8$,
these represent negligeable corrections to the above estimates.

Because the expression for the chiral condensate 
involves the square of the gap, which is only known up to a factor 
$b_0^\prime$ assumed to be ${\cal O}(1)$ \cite{Son:1999,Schafer:1999,Pisarski:1999,Brown:1999aq,Brown:1999yd}, 
there is some  theoretical uncertainty in the meson masses at low densities
and the values quoted are at most indicative. 
At this order, this uncertainty can  be 
cancelled by considering the  mass over gap ratios. These ratios   are
{\em independent} of the gap  and, moreover,  tell
us whether the mesons can exist as low energy excitations in the
CFL phase at chemical potential $\mu$.
The ratios of the charged and neutral meson masses over two times the gap are shown in Figs. 4 and 5. 
For the strange particles, the ratio is larger than one or very close to the
instability threshold unless $\mu > 1$ GeV,  
which suggests that these particles do not exist as stable low
energy excitations at low densities.


%

\section{Conclusions}


We have considered  the   effective lagrangian  for the Goldstone modes in
the Color-Flavor-Locking phase
of QCD at high densities, up to $E^4$ in a low energy expansion 
$E \lsim 2 \Delta$. 
A tentative analysis of the meson mass 
pattern that emerges from this lagrangian, including instanton effects,
suggests that the kaons and the pure strange  neutral meson may be absent from
the spectrum for moderate densities.
Their masses are significantly larger than 
two times the gap for $\mu \lsim 1$ GeV. For the pions,
the situation is  much better as there their masses are significantly smaller than the instability threshold for all densities.

Obviously, there is room for improvements. Our approach has been far from systematic and we have made
some {\em ad hoc} assumptions. Although the effects are 
likely to be small at moderate densities, it is of much interest to 
gain a better understanding of the quadratic quark mass 
effects~\cite{Hong:1999ei,Rho:2000ww}.  
A calculation of the contribution of instantons to the mass of $\eta_0$ in the chiral limit  would also prove to be most useful. If
a  spontaneously broken, approximate $U(2)_L \times U(2)_R$ flavor
symmetry  survives in the CFL phase of QCD at moderate densities,
it could also  be interesting to determine the
potential for the $\eta_0$,  $V_0(X)$. The effective theory 
for the low energy excitations should be very much analogous to the large $N_c$
lagrangian of Di Vecchia and Veneziano~\cite{Veneziano} and Witten~\cite{Witten:1980}.

\section*{Acknowledgements}

We are grateful to Deog Ki Hong, Stephen Hsu, Thomas Sch\" afer, Rob Pisarski and Dirk Rischke for useful comments.




\newpage

\begin{figure}[h]

\begin{center}

\epsfig{file=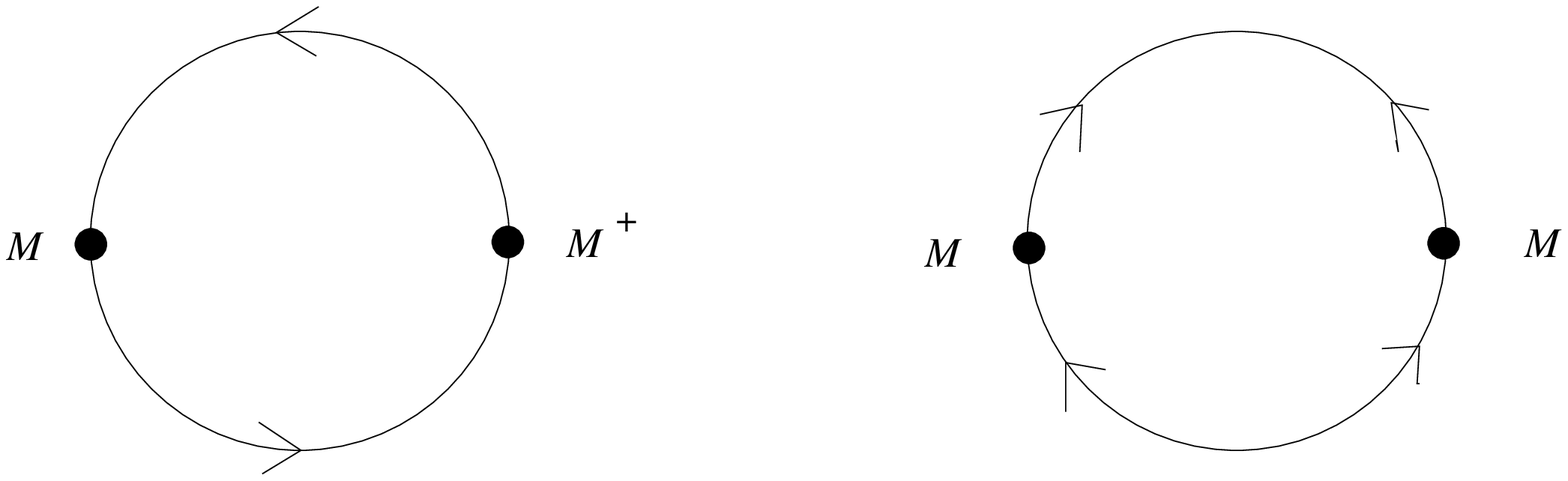,height=5cm}

\caption{Quark loops in the microscopic theory. ${\cal M}$ (${\cal M}^\dagger$) stands for one insertion of the (conjugate) quark mass matrix. The {\em lhs} diagram is like in vacuum. The {\em rhs} diagram arises only in presence of a diquark condensate.}

\end{center}

\end{figure}

\begin{figure}[h]

\begin{center}

\epsfig{file=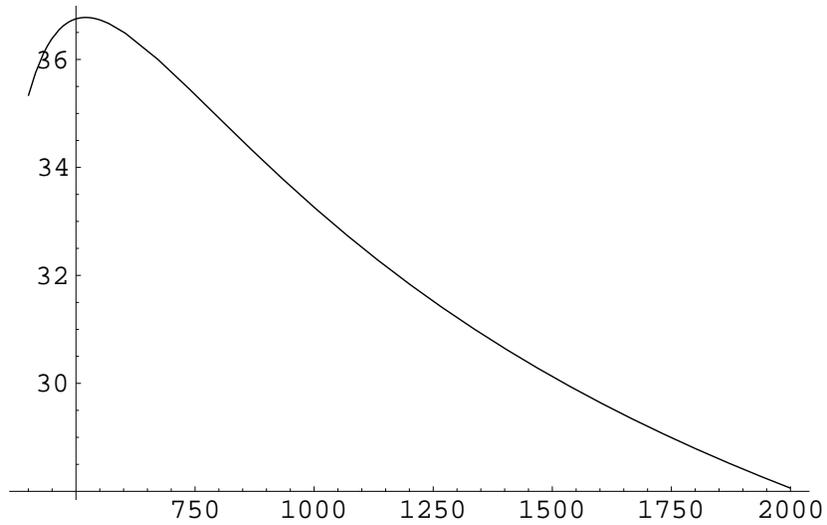,height=7cm}

\caption{Color gap (in MeV) as a function of the quark chemical potential $\mu$ (in MeV) for $b_0^\prime = 1$.}

\end{center}

\end{figure}

\begin{figure}[h]

\begin{center}

\epsfig{file=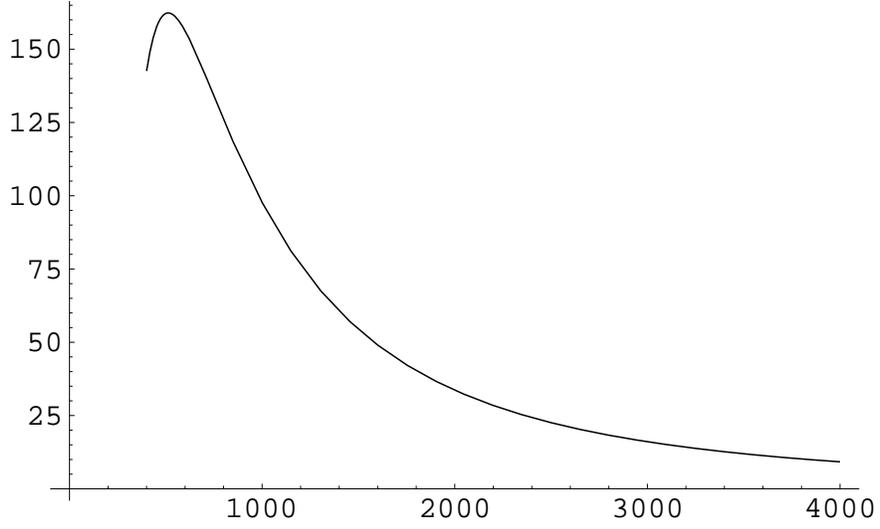,height=7cm}

\caption{Plot of   $(-\langle \bar \psi \psi\rangle)^{1/3}$ (in MeV)  as function of the quark chemical potential $\mu$ (in MeV), where $\langle \bar \psi \psi \rangle = \langle \bar u u + \bar d d + \bar s s \rangle$ is the chiral quark condensate induced by instantons.}

\end{center}

\end{figure}

\begin{figure}[h]

\begin{center}

\epsfig{file=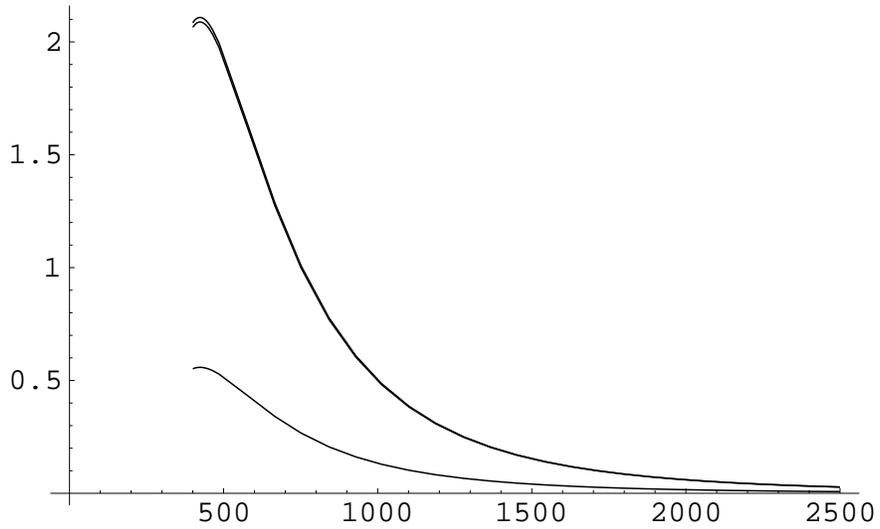,height=7cm}

\caption{Ratios of charged pions (lower curve) and kaons 
(two upper curves) masses over two times the gap as function of the quark chemical potential (in MeV).}

\end{center}

\end{figure}

\begin{figure}[h]

\begin{center}

\epsfig{file=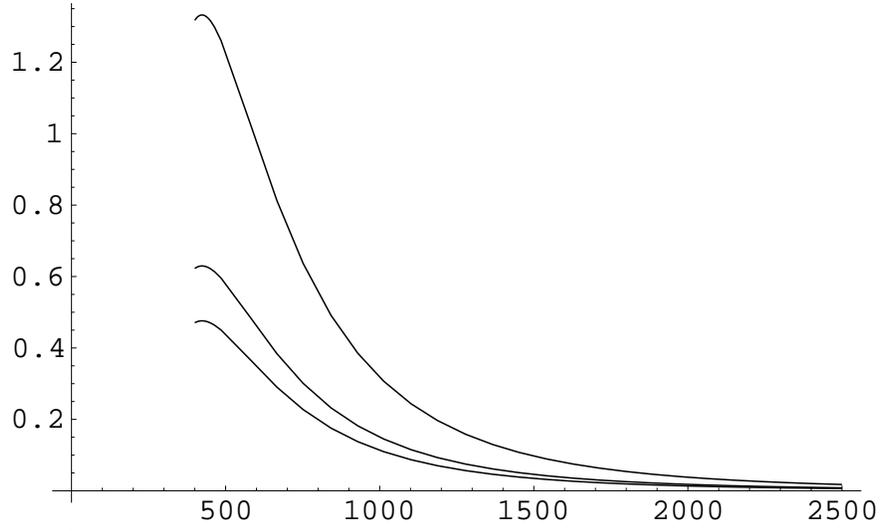,height=7cm}

\caption{Ratios of neutral meson masses ($M_{\bar u u} < M_{\bar d d} < 
M_{\bar s s}$) 
 over two times the gap as functions of the quark chemical potential (in MeV).}
\end{center}

\end{figure}













\end{document}